\documentclass[journal]{IEEEtran}
\usepackage{amsmath, amssymb, amsthm, mathtools, bbm}
\usepackage{graphicx, subfigure, caption}
\usepackage{relsize, paralist, hyperref, cite}

\newtheorem{theorem}{Theorem}
\newtheorem{lemma}[theorem]{Lemma}
\newtheorem{proposition}[theorem]{Proposition}
\newtheorem{corollary}[theorem]{Corollary}

\newtheorem{remark}[theorem]{Remark}

\newtheorem{example}{Example}

\newcommand{ \C }{ \bs{C} }
\newcommand{ \CC }{ \bs{\bar{C}} }
\newcommand{ \da }{ d_{\textnormal{a}} }
\newcommand{ \ds }{ d_{\textnormal{s}} }
\newcommand{ \Ma }{ M^{\textnormal{a}} }
\newcommand{ \Ms }{ M^{\textnormal{s}} }
\newcommand{ \tleft }{ t_{\sml{\leftarrow}} }
\newcommand{ \tright }{ t_{\sml{\rightarrow}} }

\newcommand{ \myx }{ \bs{x} }
\newcommand{ \myy }{ \bs{y} }
\newcommand{ \myS }{ \bs{S} }
\newcommand{ \myxx }{ \bs{\bar{x}} }
\newcommand{ \myyy }{ \bs{\bar{y}} }
\newcommand{ \mySS }{ \bs{\bar{S}} }
\newcommand{ \sml }[1]{ \mathsmaller{#1} }
\newcommand{ \bs }[1]{ \boldsymbol{#1} }
\newcommand{ \ceil }[1]{ \lceil #1 \rceil }
\newcommand{ \floor }[1]{ \lfloor #1 \rfloor }
\newcommand{ \simexp }{ \ \dot{\sim} \ }
\newcommand{ \weight }{ \operatorname{wt} }
\newcommand{ \myqed }{ \hfill $\blacktriangle$ }

\newcommand{ \defeq }{ \coloneqq }

\newcommand{ \deriv }[1]{ \frac{\partial}{\partial #1}}
\newcommand{ \dderiv }[1]{ \frac{\partial^2}{(\partial #1)^2}}

\begin{document}

\title{Runlength-Limited Sequences and Shift-Correcting Codes: Asymptotic Analysis}

\author{
        Mladen~Kova\v{c}evi\'c 
\thanks{Date: March 13, 2019.}%
\thanks{The author was with the Department of Electrical \& Computer Engineering,
        National University of Singapore, Singapore 117583.
				He is now with the BioSense Institute, University of Novi Sad,
				21000 Novi Sad, Serbia (email: kmladen@uns.ac.rs).}%
\thanks{This work was supported by the Singapore Ministry of Education
        (grant no. R-263-000-B61-112) and by the European Commission
				(H2020 Antares project, ref. no. 739570).}%
       }%



\maketitle

\begin{abstract}
This work is motivated by the problem of error correction in bit-shift channels
with the so-called $ \bs{(d,k)} $ input constraints (where successive $ \bs{1} $'s
are required to be separated by at least $ \bs{d} $ and at most $ \bs{k} $ zeros,
$ \bs{0 \leq d < k \leq \infty} $).
Bounds on the size of optimal $ \bs{(d,k)} $-constrained codes correcting a fixed
number of bit-shifts are derived, with a focus on their asymptotic behavior in the
large block-length limit.
The upper bound is obtained by a packing argument, while the lower bound follows
from a construction based on a family of integer lattices.
Several properties of $ \bs{(d, k)} $-constrained sequences that may be of
independent interest are established as well; in particular, the exponential
growth-rate of the number of $ \bs{(d, k)} $-constrained constant-weight sequences
is characterized.
The results are relevant for magnetic and optical information storage systems,
reader-to-tag RFID channels, and other communication models where bit-shift
errors are dominant and where $ \bs{(d, k)} $-constrained sequences are used
for modulation.%
\end{abstract}


\section{Introduction}

\IEEEPARstart{S}{hift and timing} errors are a dominant type of noise in several
communication and information storage scenarios, examples of which include magnetic
and optical recording devices \cite{hilden, shamai+zehavi},
inductively coupled channels such as the reader-to-tag RFID channel \cite{rosnes},
parallel asynchronous communications \cite{engelberg+keren},
various types of timing channels \cite{anantharam+verdu, kovacevic+popovski, nakano}, etc.
Designing codes that are able to correct these types of errors and studying their
fundamental limits is important for all these applications, in addition to being
an interesting theoretical challenge.
The problem is further complicated by the fact that, in many of the mentioned
applications, particularly magnetic, optical, and emerging DNA storage systems,
the codewords are required to satisfy modulation constraints that are introduced to
alleviate inter-symbol interference and other impairing effects
\cite{immink, immink1, immink2, immink+cai, siegel, siegel+wolf}.
Perhaps the best-known example of such constraints are runlength constraints where
a minimum and a maximum number of zero symbols between two consecutive non-zero
symbols is specified.
Motivated by these communication settings, we study the error correction problem
for channels with shift and timing errors, and with runlength input constraints.
The precise channel model we have in mind and our contributions are described in the
following two subsections.

\subsection{The Channel Model}
\label{sec:model}

Fix $ d, k \in \mathbb{Z} \!\cup\! \{\infty\} $ with $ 0 \leq d < k \leq \infty $.
We assume that the channel inputs are binary strings of length $ n $ that are
composed of blocks from the set $ \{0^d 1, \ldots, 0^k 1\} $, where $ 0^j $
is a string of $ j $ zeros.
In other words, the set of inputs is
\begin{equation}
\label{eq:dk}
  \myS_{d,k}(n) \defeq \big\{0, 1\big\}^n \cap \big\{0^d 1, \ldots, 0^k 1\big\}^* .
\end{equation}
(Here $ A^* $ is the usual notation for the set $ \bigcup_{i=0}^\infty A^i $.)
For a string $ \myx = x_1 \cdots x_n \in \{0, 1\}^n $, denote by
$ \weight(\myx) = \sum_{i=1}^n x_i $ its Hamming weight.
The set of all channel inputs of weight $ W $ is denoted by:
\begin{equation}
\label{eq:dkW}
  \myS_{d,k}(n, W) \defeq \Big\{ \myx \in \myS_{d,k}(n) : \weight(\myx) = W \Big\} .
\end{equation}
In other words, $ \myS_{d,k}(n, W) $ contains all strings of length $ n $ that
are composed of exactly $ W $ blocks from the set $ \{0^d 1, \ldots, 0^k 1\} $.

The definition \eqref{eq:dk} ensures that
\begin{inparaenum}
\item[(a)]
consecutive $ 1 $'s in any input string are separated by at least $ d $ and at
most $ k $ zeros, and
\item[(b)]
every input string starts with a string of zeros of length
$ j $, $ d \leq j \leq k $, and ends with a $ 1 $.
\end{inparaenum}
The property (a) is the defining property of the so-called $ (d,k) $-constrained
sequences.
However, the boundary conditions (b) are not universally adopted in the literature
\cite{immink}.
We shall nevertheless find it convenient to work under definition \eqref{eq:dk},
as in, e.g., \cite{krachkovsky};
it will be evident later on that different boundary conditions would not affect
the analysis in any significant way.

For a given input string $ \myx \in \myS_{d,k}(n,W) $, the channel outputs another
binary string $ \myy $ of length $ n $ and weight $ W $.\linebreak
We think of $ 1 $'s in $ \myx $ as being ``shifted'' in the channel, each for a
number of positions to the left or to the right of its original position, thus
producing $ \myy $ at the output.
We say that $ t $ bit-shifts have occurred in the channel if
$ \sum_{i=1}^W |\bar{x}_i - \bar{y}_i| = t $, where $ \bar{x}_i $ (resp. $ \bar{y}_i $)
is the position of the $ i $'th $ 1 $ in $ \myx $ (resp. $ \myy $), $ 1 \leq i \leq W $;
see Example \ref{exmpl} for an illustration.

\begin{example}
\label{exmpl}
\textnormal{
Consider an input string $ \myx \in \myS_{2,4}(15,4) $ and the corresponding
output string $ \myy $:
\begin{equation}
\label{eq:xyshift}
\begin{aligned}
  \myx \ &= \ 0 \ 0 \ 1 \ 0 \ 0 \ 0 \ 0 \ 1 \ 0 \ 0 \ 1 \ 0 \ 0 \ 0 \ 1   \\
  \myy \ &= \ 0 \ 1 \ 0 \ 0 \ 0 \ 1 \ 0 \ 0 \ 0 \ 0 \ 0 \ 1 \ 0 \ 0 \ 1 .
\end{aligned}
\end{equation}
We think of $ \myy $ as obtained from $ \myx $ by shifting the first $ 1 $
in $ \myx $ one position to the left, the second $ 1 $ two positions to the left,
and the third $ 1 $ one position to the right.
We then say that the total number of bit-shifts that occurred in the channel is
$ t = 1 + 2 + 1 = 4 $.
\myqed
}
\end{example}

Note that the output string $ \myy $ may in general violate the $ (d,k) $-constraints.

\subsection{Main Results}

Our main object of study in the present paper are error-correcting codes for
the above-described channel model.
In particular, we shall derive explicit bounds on the cardinality of optimal
$ (d, k) $-constrained codes correcting $ t $ shifts, with a focus on their
asymptotic form in the regime of growing block-length ($ n \to \infty $).
Despite a sizable body of literature on the bit-shift channel and related models%
\footnote{See, e.g., \cite{abdelghaffar+weber, bours, ferreira+lin, hilden, klove2,
kolesnik+krachkovsky, kuznetsov+hanvinck, levenshtein+hanvinck, rosnes, roth+siegel,
shamai+zehavi, ytrehus, ytrehus2}.},
such bounds, to the best of our knowledge, have not been obtained before%
\footnote{The only bounds appearing in the literature that we are aware of
are those in \cite{abdelghaffar+weber} (and \cite{ytrehus} for $ t = 1 $)
for the symmetric case; see Section \ref{sec:main_d1} ahead.
However, these bounds are not explicit and are difficult to compare to ours.
We also mention here the work \cite{kolesnik+krachkovsky}, where a lower bound
was obtained for a different asymptotic regime where $ t \sim \tau n $,
and \cite{krachkovsky}, where bounds on codes correcting \emph{all} bit-shift
errors of maximum magnitude $ s $ (zero-error codes) were derived and shown
to be tight in some cases.},
even for a single bit-shift ($ t = 1 $).

We consider two scenarios.
The first one, analyzed in Section \ref{sec:main_da}, corresponds to the situation
where shifts to the right (right-shifts) and shifts to the left (left-shifts) are
treated independently, and separate requirements on their correctability are imposed.
More precisely, codes are in this case required to have the capability of correcting
$ \tright $ right-shifts and $ \tleft $ left-shifts, for given $ \tright $ and $ \tleft $.
The second scenario, analyzed in Section \ref{sec:main_d1}, corresponds to the
situation where right-shifts and left-shifts are treated in a symmetric way, i.e.,
where codes are required to have the capability of correcting $ t $ shifts, regardless
of the direction of each individual shift.
In both cases, a metric appropriate for characterizing the error-correcting
capability of a code is given.

In Section \ref{sec:space} we state several properties of the code space
$ \myS_{d,k}(n) $ that are needed to derive the bounds in Sections \ref{sec:main_da}
and \ref{sec:main_d1} but are also of independent interest.
In particular, we determine the capacity of the noiseless channel with
$ (d, k) $-constrained \emph{constant-weight}%
\footnote{Constant-weight $ (d, k) $-constrained sequences have been studied
previously in several works \cite{kurmaev2, kurmaev}.}
inputs, i.e., the exponential
growth-rate of $ |\myS_{d,k}(n,wn)| $ as $ n \to \infty $.

\subsection{Notation}

$ \log $ denotes the base-$ 2 $ logarithm.

If we write $ \sum_i s_i $, it is understood that $ i $ ranges over all
possible values, which will be clear from the context.

We adopt the following asymptotic notation: for any two non-negative real
sequences $ (a_n) $ and $ (b_n) $,
\begin{itemize}
\item
$ a_n \sim b_n $ means $ \lim_{n \to \infty} \frac{a_n}{b_n} = 1 $;
\item
$ a_n \gtrsim b_n $ means $ \liminf_{n \to \infty} \frac{a_n}{b_n} \geq 1 $;
\item
$ a_n \simexp b_n $ means $ \log a_n  \sim  \log b_n $
(i.e., the \emph{exponents} of $ a_n $ and $ b_n $ have the same asymptotic behavior);
\item
$ a_n = {\mathcal O}(b_n) $ means $ \limsup_{n \to \infty} \frac{a_n}{b_n} < \infty $;
\item
$ a_n = o(b_n) $ means $ \lim_{n \to \infty} \frac{a_n}{b_n} = 0 $.
\end{itemize}

\section{The Space of $ (d,k) $-Constrained Sequences}
\label{sec:space}

In this section we demonstrate some properties of the set $ \myS_{d,k}(n) $
that will be used in the derivations to follow but are also of potential interest
in other applications.
We also describe another representation of this space that is equivalent to the
one given in \eqref{eq:dk}--\eqref{eq:dkW} but that may be preferable to it,
depending on the problem being analyzed.

\subsection{Equivalent Representation}
\label{sec:representation}

Another representation of the set of all channel inputs that is useful for analyzing
bit-shift errors is based on specifying the positions of $ 1 $'s in the input string
\cite{abdelghaffar+weber}.

For $ \myx \in \myS_{d,k}(n,W) $, we denote by $ \myxx = (\bar{x}_1, \ldots, \bar{x}_W) $
the vector indicating the positions of $ 1 $'s in the string $ \myx $, meaning that
$ \bar{x}_i $ is the position of the $ i $'th $ 1 $ in $ \myx $.
For example, for $ {\myx = 0 1 0 0 1 0 1 \in \myS_{1,3}(7,3)} $ we have $ \myxx = (2, 5, 7) $.
The mapping $ \myx \mapsto \myxx $ is clearly one-to-one.
With this correspondence in mind, we define
\begin{align}
\label{eq:space}
\nonumber
  \mySS_{d,k}(&n, W) \defeq  \Big\{ \myxx \in \mathbb{Z}^W  : 0 < \bar{x}_1 < \cdots < \bar{x}_W = n,   \\
  &d + 1 \leq \bar{x}_{i} - \bar{x}_{i-1} \leq k + 1  \;\; \text{for} \;\; i = 1, \ldots, W \Big\} ,
\end{align}
where it is understood that $ \bar{x}_0 = 0 $.
Hence, $ \mySS_{d,k}(n, W) $ is just a different representation of the set of
all channel inputs of length $ n $ and weight $ W $, namely $ \myS_{d,k}(n, W) $,
and $ \mySS_{d,k}(n) \defeq \bigcup_{W} \mySS_{d,k}(n, W) $ is the corresponding
representation of the set of all channel inputs of length $ n $, namely $ \myS_{d,k}(n) $.
Note that $ \mySS_{d,k}(n, W) $ is a $ (W-1) $-dimensional subset of $ \mathbb{Z}^W $
because the position of the last $ 1 $ in every input string is fixed to $ n $ by
our convention \eqref{eq:dk}, i.e.,
\begin{equation}
\label{eq:dimW-1}
  \mySS_{d,k}(n)  \subset \mathbb{Z}^{W-1}\!\times\!\{n\} .
\end{equation}

The space $ \mySS_{d,k}(n) $, or $ \mySS_{d,k}(n, W) $ in the constant-weight
case, seems to be more convenient for describing constructions of codes for the
bit-shift channel.

\subsection{Combinatorial Description and Asymptotics}

Define
\begin{subequations}
\begin{alignat}{3}
  &S_{d,k}(n)    && \defeq  | \myS_{d,k}(n) |    && = | \mySS_{d,k}(n) | ,  \\
  &S_{d,k}(n,W)  && \defeq  | \myS_{d,k}(n, W) | && = | \mySS_{d,k}(n, W) | ,
\end{alignat}
\end{subequations}
and note that
\begin{equation}
\label{eq:Wdecomp}
  S_{d,k}(n) = \sum_{W=0}^n S_{d,k}(n, W) .
\end{equation}
We see from the definition of the code space \eqref{eq:dk} that $ S_{d,k}(n) $
is in fact the number of \emph{compositions}%
\footnote{A composition of an integer $ n $ is a tuple of positive integers (called
parts) summing to $ n $ \cite{stanley}.
For a study of compositions with parts restricted to a subset of $ \mathbb{N} $
see, e.g., \cite{heubach+mansour}.}
of the integer $ n $ with parts restricted to the set $ \{d+1, \ldots, k+1\} $.
This number can be expressed in the recursive form:
\begin{equation}
\label{eq:Sdkrec}
  S_{d,k}(n) = \sum_{i=d+1}^{k+1}  S_{d,k}(n-i)
\end{equation}
with initial conditions $ S_{d,k}(0) = 1 $ and $ S_{d,k}(n) = 0 $ for $ n < 0 $.
As is well-known \cite{tang+bahl}, this implies that $ S_{d,k}(n) \sim c \rho^{- n} $,
where $ \rho $ is the unique positive solution%
\footnote{That the positive solution is unique can be seen from the fact that
the function $ \sum_{i=d+1}^{k+1}  x^{i} $ is monotonically increasing from zero
to infinity over the half-line $ x \geq 0 $.
It is also easy to see that this solution lies in the range $ (0,1) $.}
to the characteristic equation $ \sum_{i=d+1}^{k+1}  x^{i} = 1 $, and the constant
$ c $ can be obtained from the recurrence \eqref{eq:Sdkrec} and its initial conditions.

Similarly, $ S_{d,k}(n,W) $ is the number of compositions of the integer $ n $
having exactly $ W $ parts, each part belonging to the set $ \{d+1, \ldots, k+1\} $.
This quantity can be expressed in the recursive form \cite{heubach+mansour}:
\begin{equation}
\label{eq:SdkWrec}
  S_{d,k}(n, W) = \sum_{i=d+1}^{k+1}  S_{d,k}(n-i, W-1)
\end{equation}
with initial conditions $ S_{d,k}(0, 0) = 1 $, $ S_{d,k}(n, 0) = 0 $ for $ n \neq 0 $,
and $ S_{d,k}(n, W) = 0 $ for $ n < 0 $.
For $ k = \infty $ one can obtain an explicit solution to \eqref{eq:SdkWrec}
by directly counting the compositions of $ n $ with parts in $ \{d+1, d+2, \ldots\} $:
\begin{equation}
\label{eq:SdkWrecursive1}
  S_{d,\infty}(n, W) = \binom{ n - 1 - Wd }{ W - 1 } .
\end{equation}
In particular, $ S_{0,\infty}(n, W) = \binom{ n - 1 }{ W - 1 } $ and
$ S_{0,\infty}(n) = 2^{n-1} $ (recall that the last bit of every sequence is
fixed to $ 1 $ by our convention \eqref{eq:dk}).
In the following lemma we characterize the asymptotic behavior of $ S_{d,k}(n, W) $
as $ n \to \infty $ and $ {W = w n} $, for fixed
$ w \in \big[ \frac{1}{k+1}, \frac{1}{d+1} \big] $.
For simplicity, we write $ w n $ instead of, e.g., $ \floor{w n} $,
ignoring the fact that the former is not necessarily an integer.

For $ w \in \big( \frac{1}{k+1}, \frac{1}{d+1} \big) $ define the function
\begin{equation}
\label{eq:sigma}
  \sigma_{d,k}(w)  \defeq  w \log\!\sum_{i=d+1}^{k+1} \rho_w^{i-\frac{1}{w}} ,
\end{equation}
where $ \rho_w $ is the unique positive solution to
$ \sum_{i=d+1}^{k+1} \big(i - \frac{1}{w}\big) x^{i} = 0 $.
Additionally, let $ \sigma_{d,k}\big(\frac{1}{k+1}\big) = \sigma_{d,k}\big(\frac{1}{d+1}\big) = 0 $.

\begin{lemma}
\label{thm:expW}
\begin{itemize}
\item[(a)]
For any fixed $ w \in \big[ \frac{1}{k+1}, \frac{1}{d+1} \big] $,
\begin{equation}
\label{eq:expW}
  \lim_{n \to \infty} \frac{1}{n} \log S_{d,k}(n, wn)  =  \sigma_{d,k}(w) .
\end{equation}
\item[(b)]
The exponent $ \sigma_{d,k}(w) $ is a continuous, strictly concave function of $ w $.
It attains its maximal value at
\begin{equation}
  w^* = \left(\sum_{i=d+1}^{k+1} i \rho^{i} \right)^{-1}
\end{equation}
and this value is $ \sigma_{d,k}(w^*) = - \log \rho $,
where $ \rho $ is the unique positive solution to $ \sum_{i=d+1}^{k+1}  x^{i} = 1 $.
\end{itemize}
\end{lemma}

\begin{IEEEproof}
  Part \emph{(a)} of the claim, referring to the asymptotics of
$ S_{d,k}(n, wn) $ as $ n \to \infty $, follows from the known results in
analytic combinatorics \cite{pemantle+wilson}.
Namely, the generating function of the bivariate sequence $ (S_{d,k}(n, W))_{n,W} $
is obtained from \eqref{eq:SdkWrec} as:
\begin{equation}
\label{eq:gen}
\begin{aligned}
  F_{d,k}(x,y)  \defeq  \phantom{}
	                  &\sum_{n=0}^{\infty} \sum_{W=0}^{\infty}  S_{d,k}(n, W) x^n y^W  \\
                =       \phantom{}
						        &\frac{1}{1 - y (x^{d+1} + \cdots + x^{k+1})} ,
\end{aligned}
\end{equation}
wherefrom one verifies that this sequence is a Riordan array%
\footnote{Bivariate sequences with generating functions of the form $ \frac{\phi(x)}{1 - y v(x)} $
are called (generalized) Riordan arrays; see \cite[Sec.\ 12.2]{pemantle+wilson}.}
and satisfies the conditions of \cite[Thm 12.2.2]{pemantle+wilson}.
We then conclude from \cite[Thm 12.2.2]{pemantle+wilson} that
\begin{equation}
\label{eq:expW2}
  S_{d,k}(n, wn)  \simexp  v(\rho_w)^{w n} \rho_w^{-n}
                  =       \left(\sum_{i=d+1}^{k+1}  \rho_w^{i-\frac{1}{w}} \right)^{w n} ,
\end{equation}
where $ v(x) \defeq \sum_{i=d+1}^{k+1} x^i $ is the polynomial appearing
in the denominator of the generating function \eqref{eq:gen}, and $ \rho_w $
is the unique positive solution to $ \frac{x \deriv{x} v(x)}{v(x)} = \frac{1}{w} $.
This proves \eqref{eq:expW}.

Part \emph{(b)} of the claim is obtained by carefully analyzing the involved
functions.
The root $ \rho_w $ is a function of the relative weight $ w $ and is implicitly
defined by $ \sum_{i=d+1}^{k+1} (w i - 1) \rho_w^{i} = 0 $,  $ \rho_w > 0 $.
Differentiating this equation w.r.t.\ $ w $ we find that $ \deriv{w} \rho_w < 0 $.
Also, we have
$ \deriv{w} \sigma_{d,k}(w) = \log \sum_{i=d+1}^{k+1} \rho_w^{i} $
and $ \dderiv{w} \sigma_{d,k}(w) = \log e \cdot \frac{\deriv{w} \rho_w}{w \rho_w} < 0 $,
implying that $ \sigma_{d,k}(w) $ is concave.
The weight $ w^* $ that maximizes the exponent $ \sigma_{d,k}(w) $ is the one
for which $ \deriv{w} \sigma_{d,k}(w) = 0 $, i.e.,
$ v(\rho_w) = \sum_{i=d+1}^{k+1} \rho_w^{i} = 1 $.
\end{IEEEproof}
\vspace{2mm}

It is illustrative to specialize Lemma 1 to $ k = \infty $ because this case
admits an explicit solution.
After a simple calculation we get $ \rho_w = \frac{1-w(d+1)}{1-wd} $ and
\begin{equation}
\label{eq:expkinfty}
  \sigma_{d,\infty}(w) = (1 - w d) H\Big(\frac{w}{1 - w d}\Big) ,
\end{equation}
where $ H(\cdot) $ is the binary entropy function, which can also be found
directly from \eqref{eq:SdkWrecursive1} by using Stirling's approximation.
The exponent \eqref{eq:expkinfty} is maximized at $ w^* = \frac{1-\rho}{1+(1-\rho)d} $,
where $ \rho $ is the unique positive solution to $ 1 - x - x^{d+1} = 0 $.
Further specializing to $ d = 0 $, we recover the well-known fact that
$ \sigma_{0,\infty}(w) = H(w) $, in which case $ \rho = 1/2 $ and $ w^* = 1/2 $.

The quantity $ \sigma_{d,k}(w) $ defined in \eqref{eq:sigma} is the maximal
information rate (i.e., the capacity) that can be achieved in the \emph{noiseless}
channel with $ (d, k) $-constrained inputs of relative weight $ w $.
This is a refinement of the well-known result that states that the capacity of
the noiseless channel with $ (d, k) $-constrained inputs (but with \emph{no} weight
constraints) is $ \lim_{n \to \infty} \frac{1}{n} \log S_{d,k}(n) = - \log \rho $
\cite{tang+bahl}.
Namely, it follows from \eqref{eq:Wdecomp} and the fact that the exponential
growth-rate of $ S_{d,k}(n,w n) $ is maximized for $ w = w^* $ (Lemma~\ref{thm:expW})
that
$ S_{d,k}(n) \simexp S_{d,k}(n,w^* n) \simexp 2^{n \sigma_{d,k}(w^*)} = 2^{- n \log \rho} $.
The following claim strengthens this result; it asserts that the input strings of
weight (approximately) equal to $ w^* n $ account for most of the space $ \myS_{d,k}(n) $.
Informally, we say that the ``typical''%
\footnote{Lemmas \ref{thm:typicalw} and \ref{thm:typicaldk} can also be expressed
in the probabilistic language. Such a formulation is close in spirit to the statements
about high-probability sets and typical sequences in information theory \cite{cover+thomas}.}
{$ (d,k) $-constrained} strings have relative weight $ w^* $.

\begin{lemma}
\label{thm:typicalw}
  There exists a sublinear function%
\footnote{The function $ f $ may depend on the parameters $ d, k $ as well;
this is suppressed for notational simplicity.}
$ f(n) = o(n) $ such that, as $ n \to \infty $,
\begin{equation}
\label{eq:CWasymp2}
  \sum_{W \,:\, |W - w^*n| > f(n)} S_{d,k}(n, W)
    \;\lesssim\;   \frac{S_{d,k}(n)}{n^{\log n}} .
\end{equation}
\end{lemma}
\begin{IEEEproof}
  Recall that
\begin{inparaenum}
\item[(a)]
$ S_{d,k}(n) $ grows exponentially with exponent $ - \log \rho $,
\item[(b)]
$ S_{d,k}(n, W) $ grows exponentially with exponent $ \sigma_{d,k}(w) $ that
is uniquely maximized at $ w^* $, and
\item[(c)]
there are only linearly (in $ n $) many possible weights $ W $ (see \eqref{eq:Wdecomp}).
\end{inparaenum}
These facts together imply that, for any given $ \epsilon > 0 $, the number
of $ (d,k) $-constrained strings of weight $ W $ satisfying $ |W - w^* n| > \epsilon n $
is exponential with an exponent \emph{strictly smaller} than $ -\log \rho $.
More precisely, for every $ \epsilon > 0 $ there exists a (sufficiently small)
$ \gamma(\epsilon) > 0 $ such that, as $ n \to \infty $,
\begin{align}
\label{eq:CWasymp3}
  \sum_{W \,:\, |W - w^*n| > \epsilon n} S_{d,k}(n, W)
	  \;\lesssim\;   \frac{S_{d,k}(n)}{2^{\gamma(\epsilon) n}} .
\end{align}
This further implies that, for every $ \epsilon > 0 $ and large enough $ n $,
\begin{align}
\label{eq:CWasymp4}
	  \sum_{W \,:\, |W - w^*n| > \epsilon n} S_{d,k}(n, W)   \;<\;   \frac{S_{d,k}(n)}{n^{\log n}} .
\end{align}
Let $ n_0(\epsilon) $ be the smallest positive integer such that \eqref{eq:CWasymp4}
holds for all $ n \geq n_0(\epsilon) $.
Take an arbitrary sequence $ (\epsilon_i) $ satisfying
$ 1 = \epsilon_0 > \epsilon_1 > \epsilon_2 > \ldots $ and $ \lim_{i \to \infty} \epsilon_i = 0 $,
and define the function:
\begin{equation}
\label{eq:f}
  f(n)  \defeq  \epsilon_i n ,  \qquad   n_0(\epsilon_i) \leq n < n_0(\epsilon_{i+1}) .
\end{equation}
It is now easy to verify that \eqref{eq:CWasymp4} and \eqref{eq:f} imply
\eqref{eq:CWasymp2}.
\end{IEEEproof}
\vspace{2mm}

The function $ n^{\log n} $ in the statement of Lemma \ref{thm:typicalw} can
be replaced with an arbitrary sub-exponential function, but this choice is
sufficient for our purposes.
In particular, since
$ \frac{\rho^{-n}}{n^{\log n}} = o\big(\frac{\rho^{-n}}{n^t}\big) $
for any fixed $ t $, Lemma \ref{thm:typicalw} implies that one can, without
loss of generality, disregard the non-typical input strings in the asymptotic
analysis of optimal codes correcting $ t $ shifts.

In the sequel, we shall also need an estimate of the number of blocks $ 0^j 1 $,
for fixed $ j \in \{d, \ldots, k\} $, in typical input strings.
For the purpose of formally stating this result, denote by
$ S_{d,k}^{(j)}(n, W, \ell) $ the number of input strings consisting of $ W $
blocks from $ \{0^d 1, \ldots, 0^k 1\} $, exactly $ \ell $ of which are $ 0^j 1 $.
Equivalently, $ S_{d,k}^{(j)}(n, W, \ell) $ is the number of compositions of
the integer $ n $ having $ W $ parts, each part taking value in $ \{d+1, \ldots, k+1\} $,
and exactly $ \ell $ of the parts having value $ j + 1 $.
We then have $ S_{d,k}(n,W)  =  \sum_{\ell=0}^W  S_{d,k}^{(j)}(n, W, \ell) $,
for every $ j \in \{d, \ldots, k\} $.

\begin{lemma}
\label{thm:typicaldk}
  Fix $ j \in \{d, \ldots, k\} $ and denote
$ \lambda_j^* \defeq \rho^{j+1} w^*
              =      \rho^{j+1} \big(\sum_{i=d+1}^{k+1} i \rho^{i} \big)^{-1} $,
where $ \rho $ is the unique positive solution to $ \sum_{i=d+1}^{k+1}  x^{i} = 1 $.
Then, for every $ j \in \{d, \ldots, k\} $, as $ n \to \infty $ we have
\begin{equation}
\label{eq:expruns}
  S_{d,k}(n)  \simexp  S_{d,k}^{(j)}\big(n, w^*n, \lambda_j^* n\big) .
\end{equation}
Moreover, there exists a sublinear function $ f(n) = o(n) $ such that,
for every $ j \in \{d, \ldots, k\} $, as $ n \to \infty $ we have
\begin{equation}
\label{eq:runs}
  S_{d,k}(n) - 
   \sum_{\substack{W, \ell \, : \, |W - w^* n| \leq f(n) , \\ \hskip 8mm |\ell - \lambda_j^* n| \leq f(n)}}
     S_{d,k}^{(j)}(n, W, \ell) 
			\;\lesssim\;   \frac{S_{d,k}(n)}{n^{\log n}} .
\end{equation}
\end{lemma}

\begin{IEEEproof}
  Let us consider $ j = d $ and denote $ \lambda \defeq \lambda_d $ for simplicity;
the proof for general $ j $ is analogous.
The following relation is valid:
\begin{equation}
\label{eq:Sdkl}
  S_{d,k}^{(d)}(n, W, \ell)  =  \binom{W}{\ell}  S_{d+1,k}\big(n-\ell(d+1), W-\ell\big) 
\end{equation}
(the $ \ell $ parts of value $ d+1 $ can be distributed among the $ W $ parts in
$ \binom{W}{\ell} $ ways, and the remaining $ W - \ell $ parts, which are all from
$ \{d+2, \ldots, k+1\} $, form a composition of the number $ n - \ell (d + 1) $),
and therefore
\begin{equation}
\label{eq:SdknWj}
  S_{d,k}(n)  =  \sum_W \sum_{\ell} \binom{W}{\ell} S_{d+1,k}\big(n-\ell(d+1), W-\ell\big) .
\end{equation}
Since the sum in \eqref{eq:SdknWj} has polynomially many terms, we know that it
grows exponentially with the same exponent as one of its summands $ W = w n $,
$ \ell = \lambda n $, for some $ w \in \big(\frac{1}{k+1}, \frac{1}{d+1}\big) $
and $ \lambda \in (0,w) $.
By using Stirling's approximation and \eqref{eq:expW}, this exponent can be
expressed in the form
\begin{subequations}
\begin{equation}
\label{eq:expWj}
\begin{aligned}
  \lim_{n \to \infty}  &\frac{1}{n} \log\!\Bigg[ \binom{w n}{\lambda n}  S_{d+1,k}\big(n - \lambda n (d+1), (w - \lambda) n\big) \Bigg]   \\
	 &= \;  w H\Big(\frac{\lambda}{w}\Big) + (w - \lambda) \log \sum_{i=d+2}^{k+1}  \rho_{w,\lambda}^{i - \frac{1-\lambda(d+1)}{w - \lambda}}  ,
\end{aligned}
\end{equation}
where $ H(\cdot) $ is the binary entropy function and $ \rho_{w,\lambda} > 0 $
is a function implicitly defined by
\begin{equation}
  \sum_{i=d+2}^{k+1}  \left(i - \frac{1-\lambda(d+1)}{w - \lambda}\right) \rho_{w,\lambda}^i = 0  .
\end{equation}
\end{subequations}
By calculating the derivatives of the exponent (the function on the right-hand
side of \eqref{eq:expWj}) with respect to $ w $ and $ \lambda $, one finds
that it is \emph{uniquely} maximized for $ \lambda^* = \rho^{d+1} w^* $,
$ w^* = \big( \sum_{i=d+1}^{k+1} i \rho^i \big)^{-1} $, where
$ \sum_{i=d+1}^{k+1} \rho^i = 1 $.
This implies that
\begin{equation}
\label{eq:}
  S_{d,k}(n)  \simexp  S_{d,k}^{(d)}\big(n, w^*n, \lambda^* n\big)
\end{equation}
and proves \eqref{eq:expruns}.
It also implies that, for \emph{every} $ \epsilon > 0 $, the part of the sum in
\eqref{eq:SdknWj}
where $ W \leq (w^* - \epsilon) n $ \emph{or} $ W \geq (w^* + \epsilon) n $ \emph{or}
$ \ell \leq (\lambda^* - \epsilon) n $ \emph{or} $ \ell \geq (\lambda^* + \epsilon) n $
is exponentially smaller than the remaining part, from which we infer \eqref{eq:runs}
by a reasoning identical to that in the proof of Lemma \ref{thm:typicalw}
(see \eqref{eq:CWasymp3}--\eqref{eq:f}).
\end{IEEEproof}
\vspace{2mm}

In words, $ (d,k) $-constrained strings of length $ n $ typically contain
$ \sim \lambda_j^* n $ blocks $ 0^j 1 $, i.e., runs of zeros%
\footnote{A run of zeros is a block of contiguous zeros of maximal length, i.e.,
such that it is delimited on both sides either by a $ 1 $, or by the end of the
string.}
of length $ j $, and all the non-typical strings can be safely ignored in the
asymptotic analysis.
Notice that, since $ \sum_{i=d+1}^{k+1} \rho^i = 1 $, we have
$ \sum_{j=d}^k \lambda_j^* = w^* $, which is expected as $ w^* n $ is the
total (typical) number of blocks in the input strings of length $ n $.

For example, in the unconstrained case ($ d = 0, k = \infty $) we have
$ \rho = 1/2 $, $ w^* = 1/2 $, and $ \lambda_j^* = 2^{-(j+2)} $ for $ j \geq 0 $.
As a further illustration, numerical values of the quantities discussed in Lemmas
\ref{thm:expW} and \ref{thm:typicaldk} are listed in Table \ref{tab:typical}
for several archetypal $ (d,k) $-constraints;
see, e.g., \cite{immink, immink2, siegel+wolf} for applications of these
and other $ (d,k) $-constrained codes.

{
\renewcommand{\arraystretch}{1.3}
\begin{table}[h]
\centering
\caption{The exponential growth-rate of the number of $ (d,k) $-constrained
         strings ($ -\log\rho $), and the typical values of the relative
				 Hamming weight ($ w^* $) and the relative numbers of runs of zeros
				 of length $ j $ ($ \lambda_j^* $) in $ (d,k) $-constrained strings
         of length $ n \to \infty $ (rounded to three decimal places).}
 \begin{tabular}{ | c || c | c | c | c | c | }
   \hline
    $ (d,k) $           &  $ (0,2) $  &  $ (1,3) $  &  $ (1,7) $  &  $ (2,7) $  &  $ (2,10) $  \\
	 \hline
	  $ \rho $            &  $ 0.544 $  &  $ 0.682 $  &  $ 0.624 $  &  $ 0.699 $  &  $ 0.687 $  \\
	 \hline
	  $ -\log\rho $       &  $ 0.879 $  &  $ 0.551 $  &  $ 0.679 $  &  $ 0.517 $  &  $ 0.542 $  \\
	 \hline
	  $ w^* $             &  $ 0.618 $  &  $ 0.363 $  &  $ 0.295 $  &  $ 0.221 $  &  $ 0.205 $  \\
	 \hline
		$ \lambda_0^* $     &  $ 0.336 $  &      -      &      -      &      -      &      -      \\
	 \hline
		$ \lambda_1^* $     &  $ 0.183 $  &  $ 0.169 $  &  $ 0.115 $  &      -      &      -      \\
	 \hline
		$ \lambda_2^* $     &  $ 0.099 $  &  $ 0.115 $  &  $ 0.072 $  &  $ 0.075 $  &  $ 0.066 $  \\
	 \hline
		$ \lambda_3^* $     &      -      &  $ 0.078 $  &  $ 0.045 $  &  $ 0.053 $  &  $ 0.046 $  \\
	 \hline
		$ \lambda_4^* $     &      -      &      -      &  $ 0.028 $  &  $ 0.037 $  &  $ 0.031 $  \\
	 \hline
		$ \lambda_5^* $     &      -      &      -      &  $ 0.017 $  &  $ 0.026 $  &  $ 0.022 $  \\
	 \hline
		$ \lambda_6^* $     &      -      &      -      &  $ 0.011 $  &  $ 0.018 $  &  $ 0.015 $  \\
	 \hline
	  $ \lambda_7^* $     &      -      &      -      &  $ 0.007 $  &  $ 0.013 $  &  $ 0.010 $  \\
	 \hline
	  $ \lambda_8^* $     &      -      &      -      &      -      &      -      &  $ 0.007 $  \\
	 \hline
	  $ \lambda_9^* $     &      -      &      -      &      -      &      -      &  $ 0.005 $  \\
	 \hline
	  $ \lambda_{10}^* $  &      -      &      -      &      -      &      -      &  $ 0.003 $  \\
	 \hline
 \end{tabular}
\label{tab:typical}
\end{table}
}

\begin{remark}
\label{rem:typical}
\textnormal{
We emphasize that the typical values $ w^* $ and $ \lambda_j^* $ depend only
on the parameters $ d, k $ which specify the allowed lengths of runs of zeros
between consecutive ones, and not on the boundary conditions adopted in the
definition of $ (d,k) $-sequences;
see the discussion after \eqref{eq:dk}--\eqref{eq:dkW}.
For example, it is irrelevant whether or not one requires the last bit to be
$ 1 $, or whether one imposes additional requirements on the lengths of the
leading and trailing runs of zeros (the so-called $ d k l r $ constraints,
see \cite[Sec. 5.4]{immink1}).
This is because the exponential growth-rate of the cardinality of the space of
$ (d, k) $-sequences and constant-weight $ (d, k) $-sequences is not affected
by the boundary conditions.
\myqed
}
\end{remark}

\section{Codes Correcting Asymmetric Shifts}
\label{sec:main_da}

We now turn to the analysis of the bit-shift channel with $ (d, k) $ input
constraints.
The scenario we consider in this section is the one in which right-shifts and
left-shifts of $ 1 $'s are treated independently and separate requirements on
their correctability are imposed.
In particular, we shall derive bounds on the cardinality of optimal codes for this
setting.
We shall not attempt to optimize the bounds for every block-length $ n $;
rather, the focus is put on their asymptotic behavior as $ n \to \infty $.%

\subsection{Geometric Characterization}
\label{sec:geometry_da}

Suppose that $ \myxx \in \mySS_{d,k}(n,W) $ is the transmitted vector and
$ \bs{z} $ the corresponding received vector (see Section \ref{sec:representation}).
If the $ i $'th $ 1 $ in $ \myx $ has been shifted by $ k_i \in \mathbb{Z} $
positions in the channel, we will have $ z_i = \bar{x}_i + k_i $.
Thus, positive $ k_i $ means a right-shift and negative $ k_i $ a left-shift
by $ |k_i| $ positions.
Therefore, one can think of the bit-shift channel as an additive noise channel
with input alphabet $ \mathbb{N} \defeq \{1, 2, \ldots\} $.

In what follows, we denote by $ \bs{f}^+ \defeq \max\{ \bs{f}, \bs{0} \} $ and
$ \bs{f}^- \defeq \max\{ - \bs{f}, \bs{0} \} $ the positive and the negative
part of a vector $ \bs{f} $, so that $ \bs{f} = \bs{f}^+ - \bs{f}^- $
(here maximum is taken coordinate-wise).
The coordinates of $ \bs{f}^+ $ and $ \bs{f}^- $ are denoted $ f_i^+ $ and $ f_i^- $,
and are all non-negative by definition.

We say that a code $ \CC \subseteq \mySS_{d,k}(n) $ corrects $ \tright $
right-shifts and $ \tleft $ left-shifts if no two different codewords
$ \myxx, \myyy \in \CC $ can produce the same output after being impaired
with arbitrary patterns of $ \tright $ or fewer right-shifts and $ \tleft $
or fewer left-shifts.
In symbols, for every $ \myxx, \myyy \in \CC $, $ \myxx \neq \myyy $,
$ {\weight(\myx) = \weight(\myy) = W} $, and all noise vectors
$ \bs{f}, \bs{g} \in \mathbb{Z}^W $ with
$ { \sum_{i}  f_i^+  \leq \tright } $,
$ { \sum_{i}  f_i^-  \leq \tleft  } $,
$ { \sum_{i}  g_i^+  \leq \tright } $,
$ { \sum_{i}  g_i^-  \leq \tleft  } $, we have
$ \myxx + \bs{f} \neq \myyy + \bs{g} $.
Such a code $ \CC $ is said to be optimal if there is no code
$ \CC' \subseteq \mySS_{d,k}(n) $ which corrects $ \tright $ right-shifts
and $ \tleft $ left-shifts and satisfies $ {|\CC'| > |\CC|} $.

Consider the following metric on $ \mySS_{d,k}(n,W)	 $:
\begin{equation}
  \da(\myxx, \myyy)  \defeq
       \max\!\left\{\sum_{i = 1}^W  (\bar{x}_i - \bar{y}_i)^+ ,
                    \sum_{i = 1}^W  (\bar{x}_i - \bar{y}_i)^-
						\right\} .
\end{equation}
This distance is of importance in the theory of codes for asymmetric channels
\cite{klove} (hence the subscript `a').
For vectors of different dimensions (corresponding to strings of different weights),
$ \myxx \in \mySS_{d,k}(n,W_1) $, $ \myyy \in \mySS_{d,k}(n,W_2) $, $ W_1 \neq W_2 $,
we define $ \da(\myxx, \myyy) = \infty $.
The minimum distance of a code $ \CC \subseteq \mySS_{d,k}(n) $ with respect to the
metric $ \da $ is denoted $ \da(\CC) $.

The following proposition gives a metric characterization of shift-correcting codes
in the asymmetric setting.

\begin{proposition}
\label{thm:da}
  A code $ \CC \subseteq \mySS_{d,k}(n) $ can correct $ \tright $ right-shifts
and $ \tleft $ left-shifts if and only if $ \da(\CC) > \tright + \tleft $.
\end{proposition}
\begin{IEEEproof}
Suppose that $ \CC $ cannot correct $ \tright $ right-shifts and $ \tleft $
left-shifts, i.e., that there exist two distinct codewords $ \myxx, \myyy $
and two noise vectors
$ \bs{f}, \bs{g} $ with
$ { \sum_{i}  f_i^+  \leq \tright } $,
$ { \sum_{i}  f_i^-  \leq \tleft  } $,
$ { \sum_{i}  g_i^+  \leq \tright } $,
$ { \sum_{i}  g_i^-  \leq \tleft  } $, such that
$ \myxx + \bs{f} = \myyy + \bs{g} $, or equivalently,
$ \myxx + \bs{f}^+ + \bs{g}^- = \myyy + \bs{g}^+ + \bs{f}^- $.
This implies that
$ \da(\myxx, \myyy) \leq \max\!\big\{\sum_{i=1}^W (f_i^+ + g_i^-), \sum_{i=1}^W (g_i^+ + f_i^-)\big\} \leq \tright + \tleft $,
and so $ \da(\CC) \leq \tright + \tleft $.

The other direction is similar.
Suppose that $ \da(\myxx, \myyy) \leq {\tright + \tleft} $ for two distinct
codewords $ \myxx, \myyy \in \CC $, and define
$ \bs{f} = (\myyy - \myxx)^+ = \max\{ \myyy - \myxx, \bs{0} \} $ and
$ \bs{g} = (\myxx - \myyy)^+ = \max\{\myxx - \myyy, \bs{0} \} $.
Then $ \myxx + \bs{f} = \myyy + \bs{g} $ and $ f_i, g_i \geq 0 $,
$ \sum_{i=1}^W f_i \leq \tright + \tleft $, $ \sum_{i=1}^W g_i \leq \tright + \tleft $
(the last two inequalities together are equivalent to our assumption that
$ \da(\myxx, \myyy) \leq \tright + \tleft $).
We can then find non-negative vectors $ \bs{f}', \bs{f}'', \bs{g}', \bs{g}'' $
satisfying $ \bs{f} = \bs{f}' + \bs{f}'' $, $ \bs{g} = \bs{g}' + \bs{g}'' $,
$ { \sum_{i} f'_i \leq \tright } $, $ { \sum_{i} f''_i \leq \tleft } $,
$ { \sum_{i} g'_i \leq \tright } $, $ { \sum_{i} g''_i \leq \tleft } $,
and write $ \myxx + \bs{f}' - \bs{g}'' = \myyy + \bs{g}' - \bs{f}'' $.
This means that $ \CC $ cannot correct $ \tright $ right-shifts and $ \tleft $
left-shifts.
\end{IEEEproof}
\vspace{2mm}

Therefore, the error-correcting capability of a code $ \CC $ depends on the
parameters $ \tright $ and $ \tleft $ only through their sum.
In particular, $ \CC $ can correct $ \tright $ right-shifts and $ \tleft $
left-shifts if and only if it can correct $ \tright + \tleft $ right-shifts
(and $ 0 $ left-shifts).

Note that the metric space $ \left(\mySS_{d,k}(n,W), \da\right) $ is not uniform
in the sense that balls in this space have varying sizes, i.e., the size of a ball
of radius $ r $ depends on its center.
For that reason, when studying properties of codes in $ \left(\mySS_{d,k}(n,W), \da\right) $
it is sometimes more convenient to consider the unrestricted metric space
$ \left(\mathbb{Z}^{W-1}, \da\right) $ (see \eqref{eq:dimW-1}) where this effect
does not occur.
For example, we have the following expression for the cardinality of a ball of radius
$ r $ in $ \left(\mathbb{Z}^{m}, \da\right) $ \cite[Lem. 1]{kovacevic+tan_it}:
\begin{equation}
\label{eq:daball}
  B_{\textnormal a}(m,r) =
     \sum_{i}  \binom{m}{i} \binom{r}{i} \binom{r + m - i}{m - i} .
\end{equation}
The cardinality of an arbitrary ball of radius $ r $ in $ \left(\mySS_{d,k}(n,W), \da\right) $
is upper bounded by $ B_{\textnormal a}(W-1,r) $.

\begin{figure}
\centering
  \includegraphics[width=\columnwidth]{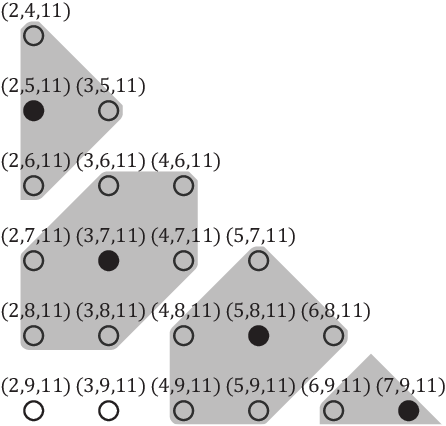}
\caption{The space $ \mySS_{1,7}(11,3) \subset \mathbb{Z}^2\!\times\!\{11\} $
representing the set of all binary strings of length $ n = 11 $ and weight
$ W = 3 $ satisfying the $ (1,7) $-constraint, and a code of minimum distance
$ \delta = 3 $ with respect to the metric $ \da $.
Codewords are depicted as black dots.
Gray regions illustrate balls of radius
$ \lfloor\frac{\delta-1}{2}\rfloor = 1 $ around the codewords.}
\label{fig:code_da}
\end{figure}

The geometric notions discussed thus far---the metric space $ \left(\mySS_{d,k}(n,W), \da\right) $
and codes in this space---are illustrated in Figure~\ref{fig:code_da}
for specific values of the parameters $ d, k, n, W $. 

\subsection{Construction and Bounds}

Denote by $ \Ma_{d,k}(n; t) $ (resp. $ \Ma_{d,k}(n, W; t) $) the cardinality
of an optimal code in $ \myS_{d,k}(n) $ (resp. $ \myS_{d,k}(n,W) $) of minimum
distance larger than $ t $ with respect to the metric $ \da $.
Since the channel does not affect the weight of the transmitted string, we know
that $ \Ma_{d,k}(n; t) = \sum_{W} \Ma_{d,k}(n, W; t) $.
The parameter $ t $ in this notation can also be understood as the sum of the
numbers of correctable right-shifts and left-shifts
(see Proposition \ref{thm:da}).

The lower bound on $ \Ma_{d,k}(n; t) $ given in Theorem \ref{thm:bounds_a_fixed}
below is obtained by constructing a family of codes in
$ \left(\mySS_{d,k}(n,W), \da\right) $, which in turn is done
by using ``good'' codes in $ \left(\mathbb{Z}^{W-1}, \da\right) $,
translating them to $ \left(\mathbb{Z}^{W-1} \!\times\! \{n\}, \da\right) $,
and then restricting to $ \mySS_{d,k}(n,W) $.
For that reason, a lower bound for codes in $ \left(\mathbb{Z}^{W-1}, \da\right) $
is given first (Lemma \ref{thm:density_a}).
This bound was essentially obtained in \cite{kovacevic+tan_it} but was not
stated there formally, so we give it here for completeness.
A few definitions are needed to state it precisely.

We say that $ {\mathcal L} \subseteq \mathbb{Z}^m $ is a sublattice of
$ \mathbb{Z}^m $ if $ ({\mathcal L}, +) $ is a subgroup of $ \big(\mathbb{Z}^m, +\big) $.
The density of $ {\mathcal L} $ in $ \mathbb{Z}^m $ is defined as
$ \mu({\mathcal L}) \defeq \big|\mathbb{Z}^m / {\mathcal L}\big|^{-1} $,
where $ \mathbb{Z}^m / {\mathcal L} $ is the quotient group of $ \mathcal L $,
and represents the average number of lattice points (from $ \mathcal L $)
per one point of the ambient space ($ \mathbb{Z}^m $).
The quantity of interest to us in the present context is the maximum density
a lattice $ {\mathcal L} \subseteq \mathbb{Z}^m $ can have when its minimum
distance is required to satisfy $ \da({\mathcal L}) > t $, namely
\begin{equation}
\label{eq:mu_a}
  \mu_{\textnormal a}(m, t)  \defeq
    \max\!\Big\{ \mu({\mathcal L}) \, : \, {\mathcal L} \subseteq \mathbb{Z}^m \, \text{a lattice}, \,
		                                \da({\mathcal L}) > t \Big\} .
\end{equation}

\begin{lemma}
\label{thm:density_a}
For every $ t \geq 1 $, as $ m \to \infty $ we have
\begin{equation}
\label{eq:density_a}
  \mu_{\textnormal a}(m, t)  \gtrsim  \frac{1}{m^{t}} .
\end{equation}
For $ t \leq 2 $ this lower bound is tight, i.e.,
\begin{equation}
\label{eq:density_a1}
	\mu_{\textnormal a}(m,1)  \sim  \frac{1}{m} , \quad
	\mu_{\textnormal a}(m,2)  \sim  \frac{1}{m^2} .
\end{equation}
\end{lemma}
\begin{IEEEproof}
It was shown in \cite[Thm 7]{kovacevic+tan_it} that every sublattice
$ {\mathcal L} \subseteq \mathbb{Z}^m $ with $ {\da({\mathcal L}) > t} $
corresponds to a Sidon set%
\footnote{A Sidon set of order $ t $ in an Abelian group $ (G, +) $ is any subset
$ \{b_0, b_1, \ldots, b_m\} \subseteq G $ with the property that all its $ t $-sums
($ b_{i_1} + \cdots + b_{i_t} $) are distinct, up to the order of the summands;
see \cite{obryant, kovacevic+tan_sidma}.}
of order $ t $ and cardinality $ m + 1 $ in an Abelian group, and vice versa.
Consequently, the largest possible density of such a sublattice can be expressed
as $ \mu_{\textnormal a}(m,t) = \frac{1}{\phi(m,t)} $, where $ \phi(m,t) $ denotes
the size of the smallest Abelian group containing a Sidon set of order $ t $ and
cardinality $ m + 1 $.
The relation \eqref{eq:density_a} then follows from the Bose--Chowla construction
of Sidon sets \cite{bose+chowla} which asserts that $ \phi(m,t) \leq m^{t} + m^{t-1} + \cdots + 1 $
when $ m $ is a prime power.

For $ t = 2 $ and $ m $ a prime power, the Bose--Chowla construction (obtained
earlier by Singer \cite{singer}) is known to be optimal, meaning that
$ \phi(m,2) = m^2 + m + 1 $.
This, together with the obvious fact that $ \phi(m,1) = m + 1 $ for every $ m $,
implies \eqref{eq:density_a1}.
\end{IEEEproof}

\begin{theorem}
\label{thm:bounds_a_fixed}
\begin{subequations}
\label{eq:bounds_a}
  For every $ t \geq 1 $ and $ d, k $ with $ 0 \leq d < k \leq \infty $,
as $ n \to \infty $ we have
\begin{align}
\label{eq:bounds_a_lower}
    \Ma_{d,k}(n; t)  &\gtrsim
		  \frac{S_{d,k}(n)}{n^{t}} \left( \sum_{i=d+1}^{k+1} i \rho^{i} \right)^{t}  ,  \\
\label{eq:bounds_a_upper}
	  \Ma_{d,k}(n; t)  &\lesssim
      \frac{S_{d,k}(n)}{n^{t}} \left( \sum_{i=d+1}^{k+1} i \rho^{i} \right)^{t} 
        \!\frac{ \ceil{t/2}! \floor{t/2}! }{ \big( (1 - \rho^{d+1}) (1 - \rho^{k+1}) \big)^{t} } ,
\end{align}
where $ \rho $ is the unique positive solution to $ \sum_{i=d+1}^{k+1}  x^{i} = 1 $.
\end{subequations}
\end{theorem}
\begin{IEEEproof}
  We first derive the lower bound \eqref{eq:bounds_a_lower}.
Consider a class of codes in $ \mySS_{d,k}(n,W) $ obtained in the following way:
Take a lattice $ {\mathcal  L} \subseteq \mathbb{Z}^{W-1} $ of minimum distance
$ \da({\mathcal L}) = t + 1 $, and let
$ {\C}_{\bs{u}} \defeq \big((\bs{u} + {\mathcal L})\!\times\!\{n\}\big) \cap \mySS_{d,k}(n,W) $
for an arbitrary $ \bs{u} \in \mathbb{Z}^{W-1} $
(here $ \bs{u} + {\mathcal L} = \{\bs{u} + \bs{x} : \bs{x} \in {\mathcal L}\} $).
Clearly, the code $ {\C}_{\bs{u}} $ has minimum distance
$ \da({\C}_{\bs{u}}) > t $. 
To give a lower bound on its cardinality notice that there are
$ \big|\mathbb{Z}^{W-1} / {\mathcal L}\big| = \mu({\mathcal L})^{-1} $
different translates $ \bs{u} + {\mathcal L} $ that are disjoint and whose
union is all of $ \mathbb{Z}^{W-1} $, so there exists at least one $ \bs{u} $
for which
$ \big|{\C}_{\bs{u}}\big| =
  \big| \big((\bs{u} + {\mathcal L})\!\times\!\{n\}\big) \cap \mySS_{d,k}(n,W) \big| \geq
  \mu({\mathcal L}) \cdot S_{d,k}(n,W) $.
This establishes the existence of a code in $ \mySS_{d,k}(n,W) $ of minimum distance
$ > t $ and cardinality $ \geq \mu_{\textnormal a}(W-1,t) \cdot S_{d,k}(n,W) $.
From Lemma \ref{thm:density_a} we then conclude that
$ \Ma_{d,k}(n, wn; t) \gtrsim \frac{S_{d,k}(n,wn)}{(w n)^{t}} $.
Finally, to get the desired lower bound on $ \Ma_{d,k}(n; t) $ write:
\begin{align}
  \Ma_{d,k}(n; t)
    \label{eq:aa}
    &=     \sum_{W}  \Ma_{d,k}(n, W; t)   \\
    &>     \sum_{W = w^*n - f(n)}^{w^*n + f(n)}  \Ma_{d,k}(n, W; t)   \\
		\label{eq:1}
    &\geq  \sum_{W = w^*n - f(n)}^{w^*n + f(n)}  \mu_{\textnormal a}(W,t) \cdot S_{d,k}(n,W)   \\
    \label{eq:2}
		&\gtrsim  \frac{ 1 }{ (w^*n + f(n))^{t} } \sum\limits_{W = w^*n - f(n)}^{w^*n + f(n)}  S_{d,k}(n,W)   \\
    \label{eq:3}
		&\sim  \frac{ S_{d,k}(n) }{ (w^*n)^{t} }  ,
\end{align}
where $ w^* $ is the optimizing weight given in Lemma \ref{thm:expW} and
$ f(n) $ is the sublinear function from Lemma \ref{thm:typicalw}.
Here \eqref{eq:1} follows from our code construction, \eqref{eq:2} follows from
Lemma \ref{thm:density_a}, and \eqref{eq:3} follows from Lemma \ref{thm:typicalw}
and the fact that $ f(n) = o(n) $.

We now turn to the derivation of the upper bound \eqref{eq:bounds_a_upper}.
Our approach is essentially a packing argument; however, due to the structure
of the code space and the fact that balls in it do not have uniform sizes, some
care is needed in making the argument work.
Let $ \C \subseteq \myS_{d,k}(n) $ be an optimal code correcting $ \floor{t/2} $
right-shifts and $ \ceil{t/2} $ left-shifts, $ |\C| = \Ma_{d,k}(n; t) $
(see Proposition \ref{thm:da}).
Consider a codeword $ \myx \in \C $ of weight $ W $, and let $ \Lambda_{j} $
(resp. $ {_j}\Lambda $) denote the number of $ 1 $'s in $ \myx $ that are
followed (resp. preceded) by exactly $ j $ zeros, and $ \Lambda_{\neq j} $
(resp. $ {_{\neq j}}\Lambda $) the number of $ 1 $'s in $ \myx $ that are
followed (resp. preceded) by a run of zeros whose length is \emph{not} $ j $.
Also, let $ {_i}\Lambda_{j} $ denote the number of $ 1 $'s in $ \myx $ that are
preceded by exactly $ i $ zeros \emph{and} followed by exactly $ j $ zeros;
$ {_i}\Lambda_{\neq j} $ the number of $ 1 $'s in $ \myx $ that are preceded by
exactly $ i $ zeros \emph{and} followed by a run of zeros whose length is \emph{not}
$ j $; and similarly for $ {_{\neq i}}\Lambda_{j} $ and $ {_{\neq i}}\Lambda_{\neq j} $.
We next show that the number of strings in $ \myS_{d,k}(n) $ that can be obtained
after $ \myx $ is impaired by $ \floor{t/2} $ right-shifts and $ \ceil{t/2} $
left-shifts is at least
\begin{equation}
\label{eq:right-shifts}
  \binom{ {_{\neq k}}\Lambda_{\neq d} - 1   }{ \floor{t/2} }
  \binom{ {_{\neq d}}\Lambda_{\neq k} - 2 t }{ \ceil{t/2}  } .
\end{equation}
To see this, first count the number of strings that can be obtained by shifting
$ \floor{t/2} $ $ 1 $'s one position to the right.
In other words, pick $ \floor{t/2} $ out of $ W $ $ 1 $'s, and shift each of them
one position to the right.
Notice that not all such choices will result in a string that belongs to the code
space $ \myS_{d,k}(n) $.
Namely, right-shifting a $ 1 $ that
\begin{inparaenum}
\item[(a)]
is preceded by exactly $ k $ zeros, or
\item[(b)]
is followed by exactly $ d $ zeros, or
\item[(c)]
is the last symbol in the string,
\end{inparaenum}
would result in either a string that violates the $ (d,k) $-constraints, or is
of length $ n + 1 $.
Excluding the $ 1 $'s satisfying (a)--(c) leaves us with at least
$ {_{\neq k}}\Lambda_{\neq d} - 1 $ $ 1 $'s to choose from, which gives the left-hand
term in \eqref{eq:right-shifts}.
The right-hand term is obtained in an analogous way by counting the number of strings
that can be obtained after picking $ \ceil{t/2} $ out of $ {_{\neq d}}\Lambda_{\neq k} - 1 $
$ 1 $'s and shifting each of them one position to the left.
The difference here is that, after choosing $ \floor{t/2} $ $ 1 $'s for the right-shifts
in the first step, we exclude additional $ 3 \floor{t/2} $ $ 1 $'s  in the second step.
Namely, if the $ i $'th $ 1 $ has been chosen for the right-shift in the first step,
then the $ (i-1) $'th, the $ i $'th, and the $ (i+1) $'th $ 1 $ are excluded in the
second step:
the $ i $'th because right-shifting and then left-shifting the same $ 1 $ would
potentially result in the same string we started with, and the $ (i-1) $'th
(resp. $ (i+1) $'th) because right-shifting the $ i $'th $ 1 $ and then left-shifting
the $ (i-1) $'th (resp. $ (i+1) $'th) could result in a run of zeros of length
$ k + 1 $ (resp. $ d - 1 $) in between these $ 1 $'s.
We are thus left with at least
$ {_{\neq d}}\Lambda_{\neq k} - 3 \floor{t/2} - 1 \geq {_{\neq d}}\Lambda_{\neq k} - 2 t $
$ 1 $'s to choose from, which yields the right-hand term in \eqref{eq:right-shifts}.
This proves our claim that the expression in \eqref{eq:right-shifts} is a lower bound
on the number of strings in $ \myS_{d,k}(n) $ that $ \myx $ can produce after being
impaired by $ \floor{t/2} $ right-shifts and $ \ceil{t/2} $ left-shifts.

We next give the asymptotic form of \eqref{eq:right-shifts}, for fixed $ t $
and $ n \to \infty $, that will be needed to conclude the proof.
We know from Lemma \ref{thm:typicaldk} that, for the ``typical'' strings in
$ \myS_{d,k}(n) $, $ {W \sim w^* n} $ and
$ {_j}\Lambda \sim \Lambda_j \sim \lambda_j^* n = \rho^{j+1} w^* n $, $ d \leq j \leq k $.
This implies that a given block $ 0^j 1 $ is preceded by a block $ 0^i 1 $ with
probability
$ \approx \frac{\lambda_i^*}{w^*} = \rho^{i+1} $, or in other words, of the
$ \Lambda_{j} $ blocks $ 0^j 1 $, a fraction of $ \approx \rho^{i+1} $ is preceded
by a block $ 0^i 1 $.
It follows that, for the typical strings in $ \myS_{d,k}(n) $,
$ {_{\neq d}}\Lambda_{\neq k}  \sim  w^* (1 - \rho^{d+1}) (1 - \rho^{k+1}) n $,
and therefore the expression \eqref{eq:right-shifts} has the following asymptotic
form:
\begin{equation}
\label{eq:shiftsasymp}
  \sim \frac{ \big(w^* (1 - \rho^{d+1}) (1 - \rho^{k+1})\big)^{t} n^{t} }{ \ceil{t/2}! \floor{t/2}! }  ,
\end{equation}
where we have used the fact that $ \binom{n}{m} \sim \frac{n^m}{m!} $ for
fixed $ m $ and $ n \to \infty $.

Finally, due to our assumption that $ \C $ corrects $ \floor{t/2} $ right-shifts
and $ \ceil{t/2} $ left-shifts, the sets of outputs that can be obtained in the
above-described way from any two different codewords $ \myx, \myy \in \C $ have
to be disjoint.
This implies that
\begin{align}
\label{eq:upper2}
  \Ma_{d,k}(n; t) \cdot
   \frac{ \big(w^* (1 - \rho^{d+1}) (1 - \rho^{k+1})\big)^{t} n^{t} }{ \ceil{t/2}! \floor{t/2}! }
    \lesssim  S_{d,k}(n)
\end{align}
and proves the upper bound in \eqref{eq:bounds_a_upper}.
We have used in \eqref{eq:upper2} the fact that in the asymptotic analysis we
can safely ignore the non-typical inputs (see Lemma \ref{thm:typicaldk}), as
we did in the derivation of the lower bound as well.
\end{IEEEproof}

\begin{corollary}
  For every $ t \geq 1 $ and $ d, k $ with $ 0 \leq d < k \leq \infty $,
as $ n \to \infty $ we have
\begin{equation}
  \log \Ma_{d,k}(n; t)  =
      - n \log \rho  -  t \log n  +  {\mathcal O}(1) ,
\end{equation}
where $ \rho $ is the unique positive solution to $ \sum_{i=d+1}^{k+1}  x^{i} = 1 $.
\end{corollary}
\begin{IEEEproof}
  The statement follows from Theorem \ref{thm:bounds_a_fixed} and the fact that
$ S_{d,k}(n) \sim c \rho^{-n} $, for a constant $ c $.
\end{IEEEproof}
\vspace{2mm}

For example, in the unconstrained case ($ d = 0, k = \infty $) we have
$ S_{0,\infty}(n) = 2^{n-1} $, $ \rho = 1/2 $, and the bounds \eqref{eq:bounds_a}
reduce to:
\begin{equation}
\label{eq:unconstrained_a}
  \frac{2^{n-1}}{n^{t}} 2^{t}
	  \lesssim
	    \Ma_{0,\infty}(n; t)
		\lesssim
  \frac{2^{n-1}}{n^{t}} 4^{t} \ceil{t/2}! \floor{t/2}! .
\end{equation}

To conclude this section, we note that the bounds \eqref{eq:bounds_a} would
continue to hold if we had adopted different boundary conditions in the
definition of $ (d, k) $-sequences \eqref{eq:dk}, because these conditions do
not affect the typical values $ w^* $ and $ \lambda_j^* $ (see Remark \ref{rem:typical}).
Of course, in that case, $ S_{d,k}(n) $ in \eqref{eq:bounds_a} would represent
the cardinality of the corresponding code space, not the cardinality of the
code space considered in this paper (see \eqref{eq:Sdkrec}).
For example, if we do not require the last bit to be $ 1 $, the bounds for the
unconstrained case would be the same as those in \eqref{eq:unconstrained_a},
with $ 2^{n-1} $ replaced by $ 2^n $.

More generally, our construction and method of deriving the bounds can be
used for the bit-shift channel with other types of input constraints---a
code would be constructed by intersecting the code space with a translated
lattice having the desired minimum distance, and the upper bound would be
derived by an analogous packing argument.
Note that, to state the resulting bounds explicitly, the typical values $ w^* $
and $ \lambda_j^* $ would first have to be determined for the constraint in
question.
Still more generally, the same approach can be used in some other weight-preserving
channels as well, such as the channel with insertions and deletions of blocks
of zeros \cite{levenshtein, kovacevic+tan_clet}.

\section{Codes Correcting Symmetric Shifts}
\label{sec:main_d1}

In this section we discuss a slightly different scenario---the one which is usually
considered in the literature on bit-shift channels---where left-shifts and right-shifts
are treated in a symmetric way.
Our object of study are codes that enable the receiver to reconstruct the transmitted
string whenever the total shift of its $ 1 $'s does not exceed a specified threshold,
regardless of the direction of each individual shift.

\subsection{Geometric Characterization}
\label{sec:geometry_d1}

Suppose that $ \myxx \in \mySS_{d,k}(n) $ is the transmitted codeword, and $ \bs{z} $
the corresponding received vector.
If the $ 1 $'s in $ \myxx $ have been shifted by $ t $ positions in total, then
$ \sum_{i=1}^W |z_i - \bar{x}_i| = t $.

We say that a code $ \CC \subseteq \mySS_{d,k}(n) $ can correct $ t $ \emph{shifts}
if no two different codewords $ \myxx, \myyy \in \CC $ can produce the same output
after being impaired with arbitrary patterns of $ t $ or fewer shifts.
In symbols, for every $ \myxx, \myyy \in \CC $, $ \myxx \neq \myyy $,
$ \weight(\myx) = \weight(\myy) = W $, and all
noise vectors $ \bs{f}, \bs{g} \in \mathbb{Z}^W $ with
$ \sum_{i=1}^W |f_i| \leq t $, $ \sum_{i=1}^W |g_i| \leq t $, we have
$ \myxx + \bs{f} \neq \myyy + \bs{g} $.
Such a code $ \CC $ is said to be optimal if there is no other code
$ \CC' \subseteq \mySS_{d,k}(n) $ correcting $ t $ shifts and such that
$ |\CC'| > |\CC| $.

\begin{figure}
\centering
  \includegraphics[width=\columnwidth]{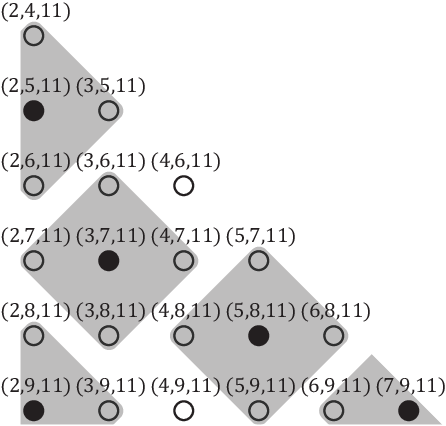}
\caption{The space $ \mySS_{1,7}(11,3) \subset \mathbb{Z}^2\!\times\!\{11\} $
representing the set of all binary strings of length $ n = 11 $ and weight
$ W = 3 $ satisfying the $ (1,7) $-constraint, and a code of minimum distance
$ \delta = 3 $ with respect to the metric $ \ds $.
Codewords are depicted as black dots.
Gray regions illustrate balls of radius $ \floor{\frac{\delta-1}{2}} = 1 $
around the codewords.}
\label{fig:code_d1}
\end{figure}

Let $ \ds $ denote the Manhattan distance on $ \mySS_{d,k}(n,W) $:
\begin{equation}
  \ds(\myxx, \myyy)
    \defeq  \sum_{i = 1}^W |\bar{x}_i - \bar{y}_i| .
\end{equation}
For vectors of different dimensions (corresponding to strings of different weights),
$ \myxx \in \mySS_{d,k}(n,W_1) $, $ \myyy \in \mySS_{d,k}(n,W_2) $, $ W_1 \neq W_2 $,
we set $ \ds(\myxx, \myyy) = \infty $.
The minimum distance of a code $ \CC \subseteq \mySS_{d,k}(n) $ with respect to
the metric $ \ds $ is denoted $ \ds(\CC) $.
The metric space $ \left(\mySS_{d,k}(n,W), \ds\right) $ and a code in this space
are illustrated in Figure \ref{fig:code_d1} for specific values of the parameters
$ d, k, n, W $.

The following proposition gives a metric characterization of shift-correcting
codes in the symmetric setting.
It was first stated in \cite[Thm 1]{kolesnik+krachkovsky} (in a more general form)
but is also implicit in some of the earlier works, e.g., \cite{abdelghaffar+weber}.

\begin{proposition}
\label{thm:d1}
  A code $ \CC \subseteq \mySS_{d,k}(n) $ can correct $ t $ shifts if and only if
$ \ds(\CC) > 2t $.
\hfill \IEEEQED
\end{proposition}

Again, the space $ \left(\mySS_{d,k}(n,W), \ds\right) $ is not uniform in the
sense that ball sizes depend on the locations of their centers.
In the unrestricted space $ \left(\mathbb{Z}^{W-1}, \ds\right) $ this effect
does not occur and we have the following expression for the cardinality of a
ball of radius $ r $ in $ \left(\mathbb{Z}^{m}, \ds\right) $ \cite{golomb+welch}:
\begin{equation}
\label{eq:d1ball}
  B_\textnormal{s}(m, r) = \sum_{i}  2^i \binom{m}{i} \binom{r}{i} .
\end{equation}
The cardinality of an arbitrary ball of radius $ r $ in $ \left(\mySS_{d,k}(n,W), \ds\right) $
is upper bounded by $ B_{\textnormal{s}}(W-1,r) $.

\subsection{Construction and Bounds}

Let $ \Ms_{d,k}(n; t) $ (resp. $ \Ms_{d,k}(n, W; t) $) denote the cardinality
of an optimal code in $ \myS_{d,k}(n) $ (resp. $ \myS_{d,k}(n,W) $) correcting
$ t $ shifts, or equivalently, having minimum distance larger than $ 2t $ with
respect to the metric $ \ds $.
Since the channel does not affect the weight of the transmitted string, we have
$ \Ms_{d,k}(n; t) = \sum_{W} \Ms_{d,k}(n, W; t) $.

In analogy with \eqref{eq:mu_a} we define the maximum density a lattice
$ {\mathcal L} \subseteq \mathbb{Z}^m $ with $ \ds({\mathcal L}) > 2t $
can have as%
\begin{equation}
  \mu_{\textnormal{s}}(m,t)  \defeq
    \max\!\Big\{ \mu({\mathcal L}) \, : \, {\mathcal L} \subseteq \mathbb{Z}^m \, \text{a lattice},\,
		                                \ds({\mathcal L}) > 2t \Big\} .
\end{equation}

\begin{lemma}
\label{thm:density_1}
For every $ t \geq 1 $, as $ m \to \infty $ we have
\begin{equation}
\label{eq:density_1}
  \mu_{\textnormal{s}}(m, t)  \gtrsim  \frac{c(t)}{m^{t}} ,
\end{equation}
where
\begin{equation}
\label{eq:ct}
  c(t) \defeq
	\begin{cases}
	\frac{1}{2t}   ,   &  t \leq 2  \\
	\frac{1}{2t+1} ,   &  t \geq 3 
  \end{cases} .
\end{equation}
For $ t = 1 $ this lower bound is tight, i.e.,
\begin{equation}
\label{eq:density_11}
	\mu_{\textnormal s}(m,1)  \sim  \frac{1}{2m} .
\end{equation}
\end{lemma}
\begin{IEEEproof}
  Consider the so-called $ A_{m-1} $ lattice defined by
$ A_{m-1} \defeq \big\{ (u_1, \ldots, u_m) \in \mathbb{Z}^m : \sum_{i=1}^m u_i = 0 \big\} $,
and let $ {\mathcal L}_0 $ be the densest sublattice of $ A_{m-1} $ satisfying
$ \ds({\mathcal L}_0) > 2t $.
By \cite[Thm 4]{kovacevic+tan_it}, the metric space $ (A_{m-1}, \ds) $ is
isometric to $ (\mathbb{Z}^{m-1}, 2 \da) $, so the density of $ {\mathcal L}_0 $
in $ A_{m-1} $ can be expressed as
$ \mu({\mathcal L}_0) \defeq | A_{m-1} / {\mathcal L}_0 |^{-1}
  = \mu_{\textnormal{a}}\big(m-1, t\big) $.
Now define a lattice $ {\mathcal L} \subseteq \mathbb{Z}^m $ by
$ {\mathcal L} \defeq \bigcup_{k \in \mathbb{Z}} \big( {\mathcal L}_0 + k (2t + 1) \bs{e}_1 \big) $,
where $ \bs{e}_1 = (1, 0, \ldots, 0) $ is a unit vector in $ \mathbb{Z}^m $.
In words, $ {\mathcal L} $ comprises infinitely many translates of $ {\mathcal L}_0 $
separated by a multiple of $ 2t + 1 $.
It follows from the construction that $ \ds({\mathcal L}) = 2t + 1 $ and
$ \mu({\mathcal L}) \defeq | \mathbb{Z}^m / {\mathcal L} |^{-1}
  = \frac{1}{2t + 1} \mu({\mathcal L}_0)
  = \frac{1}{2t + 1} \mu_{\textnormal{a}}(m-1, t) $,
which, together with Lemma \ref{thm:density_a}, implies that
$ \mu_{\textnormal{s}}(m, t) \gtrsim \frac{1}{2t+1} m^{-t} $.

The lower bound just given can be improved for $ t = 1, 2 $  to
$ \mu_{\textnormal{s}}(m,t) \gtrsim \frac{1}{2t} m^{-t} $.
In fact, for $ t = 1 $ the optimal density is known exactly for every $ m $:
$ \mu_{\textnormal{s}}(m,1) = \frac{1}{2 m + 1} $.
This follows from the existence of perfect codes of radius $ t = 1 $ in
$ (\mathbb{Z}^{m}, \ds) $ \cite{golomb+welch}.
For $ t = 2 $ (or, indeed, for any $ t $), one can construct codes in
$ (\mathbb{Z}^{m}, \ds) $ by periodically extending codes in the torus
$ \mathbb{Z}_q^m $ correcting $ t = 2 $ errors in the Lee metric%
\footnote{To the best of our knowledge, no known construction of codes in the
Lee metric gives a lower bound on the density $ \mu_{\textnormal{s}}(m,t) $ better
than the one stated in \eqref{eq:density_1}, except for $ t = 1, 2 $.
For example, Berlekamp's construction \cite[Ch.\ 9]{berlekamp} (see also \cite{chiang+wolf})
gives $ \mu_{\textnormal{s}}(m,t) \gtrsim 2^{-t} m^{-t} $, and the BCH-like
construction of Roth and Siegel \cite{roth+siegel} gives
$ \mu_{\textnormal{s}}(m,t) \gtrsim p_{2t+3}^{-1} m^{-t} $,
where $ p_{2t+3} $ is the smallest prime greater than or equal to $ 2t+3 $.}
(here $ \mathbb{Z}_q \defeq \mathbb{Z} / (q \mathbb{Z}) $).
Such a periodic extension of Berlekamp's codes for the Lee metric \cite[Ch.\ 9]{berlekamp}
gives $ \mu_{\textnormal{s}}(m,2) \gtrsim \frac{1}{4} m^{-2} $.
\end{IEEEproof}

\begin{theorem}
\label{thm:bounds_1_fixed}
\begin{subequations}
\label{eq:bounds_1}
  For every $ t \geq 1 $ and $ d, k $ with $ 0 \leq d < k \leq \infty $,
as $ n \to \infty $ we have
\begin{align}
\label{eq:bounds_1_lower}
   \Ms_{d,k}(n; t)  &\gtrsim
		  \frac{ S_{d,k}(n) }{ n^{t} } \left( \sum_{i=d+1}^{k+1} i \rho^{i} \right)^{t} c(t)  ,  \\
\label{eq:bounds_1_upper}
	 \Ms_{d,k}(n; t)  &\lesssim
      \frac{ S_{d,k}(n) }{ n^{t} } \left( \sum_{i=d+1}^{k+1} i \rho^{i} \right)^{t}
        \!\frac{ t! \, 2^{-t} }{ \big( (1 - \rho^{d+1}) (1 - \rho^{k+1}) \big)^{t} } ,
\end{align}
where $ c(t) $ is defined in \eqref{eq:ct} and $ \rho $ is the unique
positive solution to $ \sum_{i=d+1}^{k+1}  x^{i} = 1 $.
\end{subequations}
\end{theorem}
\begin{IEEEproof}
  The proof is analogous to the proof of Theorem~\ref{thm:bounds_a_fixed} for
the asymmetric case;
the only difference in proving the lower bound \eqref{eq:bounds_1_lower} is
that we need to use the metric $ \ds $ instead of $ \da $ (see Proposition
\ref{thm:d1}).
Following the same steps as in \eqref{eq:aa}--\eqref{eq:3} we get
$ \Ms_{d,k}(n; t)  \gtrsim  \mu_{\textnormal{s}}(w^*n, t) \cdot S_{d,k}(n) $,
and then the result follows by applying Lemma \ref{thm:density_1}.

Now for the upper bound \eqref{eq:bounds_1_upper}.
Let $ \C \subseteq \myS_{d,k}(n) $ be an optimal code correcting $ t $
shifts, $ |\C| = \Ms_{d,k}(n; t) $, and consider a codeword $ \myx \in \C $.
Every pattern of $ t $ shifts that can impair $ \myx $ in the channel
consists of $ r $ right-shifts and $ t - r $ left-shifts, for some
$ r \in \{0, 1, \ldots, t\} $.
By a reasoning identical to that used in the proof of Theorem \ref{thm:bounds_a_fixed}
we then conclude that the number of strings in $ \myS_{d,k}(n) $ that can be
produced after $ \myx $ is impaired by $ t $ shifts is at least
\begin{equation}
\label{eq:shifts}
  \sum_{r=0}^t
	  \binom{ {_{\neq k}}\Lambda_{\neq d} - 1  }{ r }
    \binom{ {_{\neq d}}\Lambda_{\neq k} - 3 r - 1 }{ t - r }
\end{equation}
(see equation \eqref{eq:right-shifts} and the paragraph following it).
Recalling that
$ {_{\neq d}}\Lambda_{\neq k}  \sim  {_{\neq k}}\Lambda_{\neq d}
  \sim  w^* (1 - \rho^{d+1}) (1 - \rho^{k+1}) n $ as
$ n \to \infty $, we find the asymptotics of the expression \eqref{eq:shifts}
in the form:
\begin{align}
\label{eq:shift2sasymp}
\nonumber
  &\sim \sum_{r=0}^t
	  \frac{ \big( {_{\neq k}}\Lambda_{\neq d} \big)^r }{ r! }
    \frac{ \big( {_{\neq d}}\Lambda_{\neq k} \big)^{t-r} }{ (t - r)! } 
    \;\; \sim
    \frac{ 2^t }{ t! }  \big( {_{\neq k}}\Lambda_{\neq d} \big)^t   \\
		&\sim 
		\frac{ 2^t }{ t! }  \big( w^*  (1 - \rho^{d+1})  (1 - \rho^{k+1}) \big)^t  n^t  .
\end{align}
Since $ \C $ corrects $ t $ shifts by assumption, we must have
\begin{align}
  \Ms_{d,k}(n; t)  \cdot
     \frac{ 2^t }{ t! }  \big( w^*  (1 - \rho^{d+1})  (1 - \rho^{k+1}) \big)^t  n^t
   \lesssim  S_{d,k}(n)  ,
\end{align}
which is equivalent to \eqref{eq:bounds_1_upper}.
\end{IEEEproof}
\vspace{2mm}

\begin{corollary}
For every $ t \geq 1 $ and $ d, k $ with $ 0 \leq d < k \leq \infty $,
as $ n \to \infty $ we have
\begin{equation}
  \log \Ms_{d,k}(n; t)  =  - n \log\rho - t \log n  +  {\mathcal O}(1) ,
\end{equation}
where $ \rho $ is the unique positive solution to $ \sum_{i=d+1}^{k+1}  x^{i} = 1 $.
\hfill \IEEEQED
\end{corollary}
\vspace{2mm}

In the unconstrained case ($ d = 0, k = \infty $) we have
$ S_{0,\infty}(n) = 2^{n-1} $, $ \rho = 1/2 $, and the bounds \eqref{eq:bounds_1}
reduce to:
\begin{equation}
\label{eq:unconstrained1}
  \frac{2^{n-1}}{n^{t}} 2^t c(t)
    \lesssim
	   \Ms_{0,\infty}(n; t)
	  \lesssim
  \frac{2^{n-1}}{n^{t}} 2^t t! .
\end{equation}

Similarly as in the asymmetric case, the bounds \eqref{eq:bounds_1} would
continue to hold if we had adopted different boundary conditions in the definition
of $ (d, k) $-sequences \eqref{eq:dk}, because these conditions do not affect
the typical values $ w^* $ and $ \lambda_j^* $ (see Remark \ref{rem:typical}).
For example, if one does not fix the value of the last bit, the bounds for the
unconstrained case would be the same as those in \eqref{eq:unconstrained1},
with $ 2^{n-1} $ replaced by $ 2^n $.

\section{Concluding Remarks}
\label{sec:conclusion}

We conclude the paper with a few remarks on error models related to those we have
studied here.

In some applications it is reasonable to assume that the shifts are limited in the
sense that each $ 1 $ in an input string $ \myx $ can be shifted by at most $ s $
positions \cite{abdelghaffar+weber, krachkovsky, shamai+zehavi}.
More precisely, if $ \myxx $ is the transmitted vector and $ \bs{z} $ the received
vector, the assumption here is that $ | z_i - \bar{x}_i | \leq s $, $ {1 \leq i \leq n} $.
For such models, the lower bound on the cardinality of optimal shift-correcting codes
can possibly be improved by using the known constructions of codes for limited-magnitude
errors; see, e.g., \cite{cassuto}.
For example, in the asymmetric case with $ \tright + \tleft = 2 $, $ s = 1 $,
the lower bound in \eqref{eq:bounds_a_lower} can be improved by a factor of $ 2 $
by using a construction of codes correcting $ 2 $ asymmetric $ 1 $-limited-magnitude
errors \cite[Sec. IV.C]{cassuto}.
Note that we have implicitly used the assumption that $ s = 1 $ in our derivation
of the upper bounds \eqref{eq:bounds_a_upper} and \eqref{eq:bounds_1_upper}.
Therefore, these upper bounds are not likely to be improved in limited-shift
models by using the approach we have used.

In this context one may also be interested in codes correcting \emph{all}
possible patterns of shifts such that each shift is bounded by $ s $.
Such (zero-error) codes have been studied in several related settings:
bit-shift channels \cite{shamai+zehavi, krachkovsky}, timing channels
\cite{kovacevic+popovski, kovacevic+stojakovic+tan}, skew-tolerant parallel
asynchronous communications \cite{engelberg+keren, kovacevic}, etc.
In many cases, the optimal codes have been found and the zero-error capacity
of the corresponding channel determined.

\section*{Acknowledgment}

The author would like to thank
Vincent Y. F. Tan (NUS), for his detailed reading and helpful comments on a preliminary
version of this work;
Mehul Motani (NUS), for several discussions on a model related to the one studied
in this paper;
Anshoo Tandon (NUS), for the many helpful discussions on constrained codes and
related notions;
and the three anonymous referees, whose detailed comments and corrections have
substantially improved the manuscript.

\IEEEtriggeratref{19}


\end{document}